# Far-field subwavelength acoustic imaging by deep learning

Bakhtiyar Orazbayev and Romain Fleury[*]

Laboratory of Wave Engineering, École Polytechnique Fédérale de Lausanne (EPFL), 1015 Lausanne, Switzerland

(Dated: November 15, 2019)

Seeing and recognizing an object whose size is much smaller than the illumination wavelength is a challenging task for an observer placed in the far field, due to the diffraction limit. Recent advances in near and far field microscopy have offered several ways to overcome this limitation; however, they often use invasive markers and require intricate equipment with complicated image post-processing. On the other hand, a simple marker-free solution for high-resolution imaging may be found by exploiting resonant metamaterial lenses that can convert the subwavelength image information contained in the near-field of the object to propagating field components that can reach the far field. Unfortunately, resonant metalenses are inevitably sensitive to absorption losses, which has so far largely hindered their practical applications. Here, we solve this vexing problem and show that this limitation can be turned into an advantage when metalenses are combined with deep learning techniques. We demonstrate that combining deep learning with lossy metalenses allows recognizing and imaging largely subwavelength features directly from the far field. Our acoustic learning experiment shows that, despite being thirty times smaller than the wavelength of sound, the fine details of images can be successfully reconstructed and recognized in the far field, which is crucially enabled by the presence of absorption. We envision applications in acoustic image analysis, feature detection, object classification, or as a novel noninvasive acoustic sensing tool in biomedical applications.

DOI:

[*] romain.fleury@epfl.ch

# I. INTRODUCTION

The performance of microscopy applications is usually hindered by a fundamental rule that is difficult to break – the diffraction limit [1]. According to this principle, the ultimate far-field resolution of direct wave imaging devices is intrinsically constrained by the wavelength of operation. In past decades, tremendous advances have been made in both near field [2,3] and far field [4,5] microscopy to bend this rule and allow for imaging with subwavelength resolution, either by using near field-sensors or by introducing blinking markers and taking multiple far-field images. Such techniques are often associated with sophisticated and expensive optical setups, either relying on a time-consuming near field scan or on taking multiple far-field images of samples labeled with fluorescent molecules, typically followed by extensive image post-processing [6,7]. Besides, many biomedical applications require label-free solutions (without using any fluorescent tagging) to perform remote, nondestructive, and noninvasive investigations of objects.

The origins of this limit stem from the fact that the evanescent waves scattered from the subwavelength details of an object cannot propagate to the far field, unlike the larger image features, which inevitably limits the resolution of conventional imaging techniques. Thus, to be able to recover the sub-diffraction details of the image, one needs to recover the evanescent field components, for instance, by working directly in the near field and exploiting negative refraction [8–11]. Recently, exciting alternative approaches have been proposed to convert the evanescent field components into propagating waves by using metamaterials [8,12–17]. For instance, hyperbolic dispersion can be used to gradually convert evanescent components into a wave that can propagate in the surrounding medium and reach the far field [12,18,19]. Another interesting marker-free approach to beat the diffraction limit combines a locally-resonant metamaterial lens (metalens) and time-reversal techniques [14,20,21]. The use of superoscillations [22–26] is an alternative route based on tailoring the interference of several coherent sources to focus the probe field directly into a subwavelength spot. However, these label-free approaches face difficult challenges: while metalenses are prone to losses due to their resonant nature, superoscillations are surrounded by large side-bands and typically lead to low signal-over-noise ratios [27].

More recently, advances in machine learning provided scientists from different research fields with a unique tool to solve complex problems – deep learning [28,29]. A deep neural network (DNN) composed of multiple processing layers with non-linear modules is capable of discovering and learning the intricate structure hidden in complex data by self-adjusting the internal parameters of each of its layers. By composing a sufficient number of such layers, DNNs can learn very complex functions without human intervention, allowing for many applications in different domains of science, such as engineering, biology, medicine, quantum physics, etc. [30–33]. Recent examples of deep learning successes include medical image analysis [34], speech recognition [35], image classification [31], inverse imaging problems [36–38], and all kinds of complex analytical problems [39,40]. Moreover, in search of more efficient schemes of deep learning, several hybrid schemes were proposed that integrate the physical layers into DNN [41–44]. Inspired by such tremendous success, several deep learning approaches for microscopy applications were proposed [38,45,46], where, however, DNNs are mainly used to enhance the quality of images that are obtained with traditional methods [38,46,47], exploiting, for instance, generative adversarial networks [48,49].

In this work, we propose a combination of the modern deep learning techniques and metamaterial approaches to solve the limitations mentioned above of non-invasive subwavelength imaging and open a new path for novel applications in marker-free imaging technologies. Remarkably, we show that, in stark contrast with conventional methods, the presence of absorption losses in the metalens is crucial to enable efficient learning. By putting a purposely lossy locally-resonant metalens in the proximity of subwavelength input images and training the DNNs to reconstruct and classify them directly, we can recover details as small as $\lambda/30$ and reach a far-field experimental classification accuracy of ≈ 80%. This strategy, which is here experimentally demonstrated with an airborne audible sound, may be translated to electromagnetic waves [50].

### 1. Concept of far field subwavelength imaging.

The scheme for far-field subwavelength acoustic imaging studied in this work is illustrated in Fig. 1. We first consider a subwavelength acoustic source, shape like the digit "five." In this reference case, the signals captured by a microphone array placed in the far-field do not contain any information about the subwavelength details of the source, due to the diffraction limit (see Appendix A). In other words, regardless of the signal processing strategy used, it is not possible to image the source. Less complicated tasks, for instance, guessing digits drawn by the source with reasonable probability (a standard classification problem), are also not possible. Instead, [Fig. 1(b)], to allow for the information about the subwavelength details to reach the far-field, we insert a lossy locally resonant metalens [16,51] composed of a cluster of randomly placed subwavelength Helmholtz resonators, whose resonant modes can couple to the evanescent waves, and radiate into the far field [14]. Then, the amplitude and phase of the far field sampled by our microphone array are fed into DNNs, which is trained on many handwritten digits to be able to reconstruct and classify subwavelength images [Fig. 1(c)]. In our work, two different types of DNNs are used: a "U-net" type CNN (UCNN) [52] for the image reconstruction and a multilayer parallel CNN (PCNN) for the image classification (see Appendix C, 1 for more details). For demonstrative purposes, the input images are taken from the MNIST database of handwritten digits (70000 images with 20 × 20 pixels resolution) and downsampled to 8 × 8 pixels, as shown in Fig. 2(a). The spatial frequency domain information reveals that the square source images, with an overall dimension of $D \approx 0.1\lambda_0$, contain features with spatial frequencies $\xi_{x,y}$ up to $30/\lambda_0$, i.e. subwavelength details of size down to $\lambda_0/30$, where $\lambda_0$ is the wavelength in the surrounding medium at the operating frequency $f_0$. Our goal is to recover these high spatial frequencies using deep learning. To demonstrate this possibility in the most general way and underline the key physical ingredients for an efficient learning process, we start with a semi-analytical 2D model of the problem based on a coupled 2D dipoles method [53], which allows us to generate the necessary data for training on subwavelength image recognition.

## II. RESULTS
### 1. Semi-analytical 2D data generation based on the coupled dipole method.

In this model, the subwavelength images are drawn by driven two-dimensional dipoles in an eight by eight square array of total size $D \approx 0.1\lambda_0$ and pitch $d \approx 0.01\lambda_0$ [see Fig. 2(b)]. The response function of the dipoles is described by Lorentzian polarizability with resonance frequency $f_r$, which are coupled to each other by the 2D Green's function (see Appendix for the details of the coupled dipole model). We enforce monopolar radiation of the array by choosing the resonance frequency such that $f_r \gg f_0$, and first consider the case of the digit alone, in the absence of a metalens. To capture the spatial diversity of the source, we probe the field in four points separated by a 40° angle, placed either in the near ($0.25\lambda_0$) or far ($300\lambda_0$) field region, and at four equidistant frequencies in the range between $0.8f_0 - 1.1f_0$, thus, generating a database with 70000 samples of measured amplitudes and phases of the field (we use 60000 training samples and 10000 test samples). The results for image reconstruction and classification using the trained DNNs are summarized in Fig. 2(b). In both near and far fields, the output images [second column inset of Fig. 2(b)] hardly represent the digit "five," translating the fact that the UCNN struggles to reconstruct the test samples. This fact is confirmed in the spatial frequency domain, where UCNN reconstructs relatively well the low spatial frequencies, while the high spatial frequencies are incorrectly guessed as being much smaller, explaining the blurred shapes of the digits. Remarkably, unlike the case of image reconstruction, the classification performance of the PNN is relatively high [the last column of Fig. 2(b)], with 67.5% and 57.5% for the near and far field regions, respectively. Such accuracy is significantly higher than random guessing (10% accuracy). The relevant information about the nature of the digit that allows for partial classification is already encoded in the lowest spatial frequency components of the image, for instance in the location of its edges (consistent with the robust classification of specific digits, for example "0" and "1" shown in supplementary Fig. S4 [54]) or of its center of gravity. Therefore, with some analogy to the stimulated emission depletion technique [5], the DNN can also be used to resolve two very close sources, which, in principle, will be only limited by the accuracy of measuring the phase wavefronts.

To further improve both the reconstruction quality and the classification accuracy, we now add a metalens, in the form of $N$ randomly distributed undriven dipoles surrounding the source [see left inset of Fig. 2(c)]. Unlike the dipoles used to draw the source digits, which are driven and operated well below their resonance frequency $f_r$, the dipoles composing the metalens are undriven but resonant at $f_0$, allowing for strong multiple scattering in the lens (see details in Appendix B). From this figure, one can observe that the source digit excites modes in the resonant metalens, turning the monopole source digit into a multipolar source with more open radiation channels. It should be noted here that the number of eigenmodes in such metalens depends mostly on the number of dipoles $N$ and not on their spatial distribution (see Supplemental Material for more details). Therefore, each symbol encodes information about its subwavelength features into the way it excites different modes in the metalens, and this information is subsequently radiated into the far field. We, therefore, expect that the reconstruction quality and classification accuracy will increase with the number of modes that can be excited in the metalens at a given frequency. This fact means that, surprisingly, the introduction of losses can improve the learning efficiency by broadening the width of the metalens resonances, increasing the number of modes (see Fig. S3 in Supplemental Material [54]) that can be accessed at a given frequency (effective mode density). Such unusual dependence is in stark contrast with conventional imaging methods based on metalenses, which are typically hindered by losses.

To validate this intuition, we start by using a lossless metalens with a relatively low mode density, with only $N$ = 29 dipoles. We see in the top row of Fig. 2(c) that, while the classification accuracy of the PCNN increased from 57.5% to 74%, the image reconstruction of the digits are still of low quality (see Fig. S4 in Supplemental Material [54] for digits other than "five"). The incapability of the UCNN to accurately recover the high spatial frequency components of the subwavelength images is due to the insufficient density of states supported by the metalens, resulting in its inability to encode enough information about the subwavelength details of the image. This density of states can be improved either by increasing the number of resonators that form the metalens or by introducing absorption losses to the resonators (see

Supplemental Material [54] for quantitative calculations of the density of states for a different number of resonators and amount of added losses). In Fig. 2(c) (bottom row), we report the results obtained with *N* = 302 lossy resonators (with a collision frequency $\Gamma \approx 0.27\omega_0$). We see that not only the overall classification accuracy raised to 84%, but also that the reconstruction of the subwavelength images is now very accurate. Remarkably, while in most metalens imaging scenarios, the presence of losses inevitably degrades the imaging performance of these resonant systems, here the effect is opposite since the neural network can learn from the larger amount of information hidden in an increased number of lossy modes. The degradation of the imaging performance is mostly attributed to the fact that the high-frequency modes (with higher transverse wavevector components) have longer lifetimes (or higher quality factors) and are more influenced by dissipation losses. Therefore, other techniques for subwavelength imaging (for instance, a time-reversal technique [21]) require having discrete resonances to decode the high spatial frequency details. Unfortunately, such high resolution of modes becomes impossible with increased intrinsic losses. In contrast, in our approach, by adding losses, we increase the transferring of the subwavelength information carried by the modes with high spatial frequency details to other modes without losing such information. However, as anticipated, excessive absorption losses in a dense resonant metalens (with a high number of resonators) will result in a quick decay of modes with longer lifetimes and, therefore, will provoke a loss of the subwavelength details, see Appendix B and Fig. 8 (a), (b). Such difference may constitute a key advantage for learning methods using metalenses in the development of future acoustic or photonic applications, as losses are an integral part of any realistic wave device. The next session provides an experimental demonstration of these findings using airborne audible sound.

## 2. Experimental deep learning for subwavelength image recognition.

For the experimental demonstration of subwavelength acoustic imaging, we used an experimental setup shown in Figs. 3(a) and 3(b) (see Supplementary material for the illustration of the setup with metalens). In the experiment, the acoustic images were drawn on an 8 × 8 lattice of speakers [total size 718.24 cm$^2$, see left inset of Fig. 3(a)], each of them connected to the output's channels of a Speedgoat® real-time target machine (four IO131 modules), which allowed independent control over the voltages applied to each of the speakers (for more details see Appendix C, 3). The metamaterial lens used for enhanced image reconstruction consists of 39 subwavelength Helmholtz resonators (plastic spheres with a diameter of $D_b$ = 100 mm ≈ $\lambda_0/13$, and a neck of length 10 mm and diameter 21.5 mm, as shown in the inset of Fig. 3(c), which are randomly placed in a mesh bag, with a period roughly equal to their diameter. To define the operating frequency range, we extract scattering parameters ($S_{11}$, $S_{21}$) of a single Helmholtz resonator placed in a tube (see Appendix C for more details). The extracted scattering parameters are plotted in Fig.

3(c), from where the resonance frequency can be inferred ($f_r \approx 268$ Hz). The learning data is generated by drawing the source digit on the loudspeaker array and measuring the complex acoustic pressure at four locations and four arbitrarily chosen frequencies between $f_{min} = 220$ Hz and $f_{max} = 260$ Hz ($\lambda_{max} = 1559$ mm and $\lambda_{min} = 1319$ mm, respectively). Therefore, the overall size of the acoustic source is around $0.2\lambda$, and the digits contain features down to 30 times below the diffraction limit. Our experimental results for the subwavelength imaging in the presence and absence of metalens are summarized in Fig. 4. First, we tested the possibility of image reconstruction in the near field ($\approx 0.1\lambda_{max}$) without metalens, which, as expected in this 3D scenario, provides a proper restoration of sub-diffraction details of subwavelength images, as shown in Fig. 4(a). The trained UCNN reconstructs most of the test digits with excellent visual fidelity, although somehow blurring small details (due to a partial loss of spatial frequencies $\xi_{x,y} > 15/\lambda$, see Fig. S5), resulting in good classification accuracy (86.5%). In the far field region and the absence of metalens, the UCNN is no longer capable of resolving the details < 90 mm, resulting in the reliable reconstruction of only a few digits, such as "0" and "1" (see Fig. 4(b) and supplementary Fig. S5 [54]). Conversely, in the presence of metalens [Fig. 4(c)], the UCNN recovers the subwavelength images with an excellent visual correspondence (with correlation coefficients > 0.7). The digit classifier PCNN can successfully train on the restored images and demonstrates a good classification accuracy of 79.4%, confirming the ability of the DNN approach to recover the small subwavelength details in the far field.

### 3. Transfer learning for subwavelength imaging.

In the previous section, we demonstrated that our DNNs could restore the initial subwavelength image from the recorded amplitude-phase distributions in the far field. Here, we go one step further and demonstrate its ability to re-learn quickly on a new database, which can be much smaller than the original one. Such flexibility in the learning process is also known as transfer learning: we create a new database consisting of 600 training and 200 test samples ($\approx$ 1% of initial MNIST database) of four letters "E", "F", "L", "P" and retrain our UCNN (previously trained on the MNIST database) on this new, significantly smaller dataset. Then, we ask the neural network to classify and reconstruct unknown letters drawn in a test dataset. The experimentally reconstructed letters are shown in Fig. 5 (see Fig.S7 for more examples). The excellent visual fidelity (with correlation coefficients ≥ 0.94 between the input and reconstructed letters) demonstrates the high adaptability of the DNN approach, which becomes more efficient at learning new data types, without being limited by the diversity of the input databases.

### III. Discussion

We have experimentally demonstrated that a combination of a resonant metalens and DNNs enables the reconstruction and recognition of subwavelength acoustic images from the far field with an accuracy of ≈ 80%, which is allowed by the presence of substantial absorption losses in the metalens. This marker-free method allows for beating the diffraction limit and reconstructs source images containing details smaller than $\lambda/30$, which are crucial for accurate classification of these images, whose total size is $0.2\lambda$. Moreover, once the DNNs are trained, they significantly accelerate the subwavelength imaging process, allowing image reconstruction and recognition at high frame rates. We believe that even smaller acoustic objects can be successfully recognized by working with acoustic resonators with lower resonance frequencies, such as ones based on membranes, which can form the basis of a new form of acoustic metasurfaces dedicated to learning systems capable of non-invasive, marker-free subwavelength imaging. A potentially interesting future direction may be to explore whether a similar approach can be used to detect the presence, position, or shapes of small particles in multiple-scattering media, with potential impact in bioengineering applications. The insensitivity of the learning scheme to geometrical disorder and the non-detrimental role of absorption represents clear advantages of the method, suggesting its possible transposition to optics. Photonic subwavelength imaging may be implemented using simple CCD sensors and readily available subwavelength photonic resonators, with Lorentzian dispersion, including plasmonic particles, quantum dots, dielectric Mie resonators, or diamond vacancies.

## ACKNOWLEDGMENTS

This work was supported by the Swiss National Science Foundation (SNSF) under the Eccellenza grant No. 181232.

**APPENDIX A: ROLE OF SPATIALARRANGEMENT AND RESONANCE FREQUENCY OF SCATTERERS**

To overcome the diffraction limit without introducing markers and employing several measurements (like in STED microscopy), one needs to recover the waves that carry the subwavelength (with high spatial frequencies) details of the object. Since such waves have high transverse wavevector components and their phase velocities exceed the phase velocity of sound in free space, they are bound to the object or being of evanescent nature. Therefore, to recover such evanescent waves, they need to be converted to propagating waves. Such conversion can be performed by coupling these waves to the modes of the resonant metamaterial that have similar high transverse wavevector $k_\perp$ components, as illustrated in Fig. 6 (blue dashed line represents the dispersion relation of such resonant metamaterial). In such resonant metamaterial, composed of tightly packed multiple resonators with distances $a_d \ll \lambda_0$ and a resonance frequency $\omega_0$, the multiple scattering provides the existence of the modes (at frequencies $\omega < \omega_0$) with high transverse wavevector components that, in turn, can leak into free space and be captured in the far field. From Fig. 6, it can be also seen that it is possible to retrieve some of these high transverse wavevectors without any conversion but at a higher operating frequency $\omega_f = 10\omega_0$, i.e., outside of the subwavelength regime (shown as a semi-transparent blue region).

Thus, to reconstruct a subwavelength source (with overall dimensions $0.1\lambda_0 \times 0.1\lambda_0$) we surround it with a resonant metalens composed of periodic ($a_d = 0.005\lambda_0$) scattering particles resonating at the frequency $\omega_0$, as shown in the left inset of Fig. 7(a). The number of modes in such resonant medium depends on the number of resonators and not their spatial distribution, as we will demonstrate in the next part. Therefore, for a more striking illustrative purpose, we use a high number of resonators (300) to construct the metalens. As expected, the neural network can reconstruct the images with high fidelity by learning on complex field amplitudes in the far field (with a classification accuracy of 83.6%). For more explicit illustration, a few examples of input and reconstructed images are shown in the first and second rows of Fig. 7(a).

Next, to demonstrate that the high $k_\perp$ modes are due to the multiple scattering and do not depend on the spatial distribution of the resonators but the strong coupling between them, we introduce some random disorder to the lattice in the range $\delta r = [0, a_d]$ [Fig. 7(b)] and $\delta r = [0, 2a_d]$ [Fig. 7(c)]. As can be seen, the neural network can reconstruct images with similar accuracy (classification accuracies of 84.6% and 84%). However, increasing the resonance frequency of resonators (for instance $\omega_1 = 3\omega_0$) will modify the dispersion relation and result in a weaker coupling between them at $\omega < \omega_0$, which is shown in Fig. 6 (red dashed curve). Thus, the subwavelength details can no longer be recovered, resulting in a low classification accuracy, which is close to the case of bare digits, as shown in Fig. 7(d). This fact demonstrates the importance of multiple scattering phenomena, which is crucial for the conversion of the evanescent waves.

## APPENDIX B: INFLUENCE OF ABSORPTION LOSSES

To further demonstrate the importance of the absorption losses for the reconstruction of subwavelength images, we performed additional simulations for metalens with 300 dipoles and different values of absorption losses (by controlling the damping factor of resonators) as shown in Fig. 8(a), (b). In the presence of radiation losses only ($\Omega = 0$ and the collision frequency $\Gamma = 0.25\omega_0$) the classification accuracy does not surpass 80%. Next, we increase the accuracy by adding the absorption losses and therefore widening average resonance linewidth $\langle \Gamma_n \rangle$. It should be noted here, that the high-frequency modes (that also have higher $k_\perp$) have longer lifetimes or higher quality factors and therefore are more influenced by the dissipation losses. In other approaches, for instance, the time-reversal technique for subwavelength imaging [21], decoding of the subwavelength information requires resolving the maximum of modes (having discrete resonances), which becomes impossible with increased intrinsic losses. However, in our approach, by adding losses we first increase transferring of the subwavelength information carried by the modes with high $k_\perp$ to the lower modes, without losing such information. However, as anticipated, excessive absorption losses will result in a too-quick decay of the modes with high $k_\perp$ and therefore will provoke a loss of the subwavelength details, see Fig. 8 (a), (b). Our study shows that the optimal absorption level corresponds to resonators in which the decay rate is around 30% of their resonance frequency. In our experiment, the Q factor of the Helmholtz resonators is around 7, consistent with operation with optimal absorption level.

## APPENDIX C: METHODS

### 1. Deep neural network architectures.

The deep neural networks used in this work are based on the convolutional neural networks (CNNs), commonly applied in many visual recognition tasks, including image reconstruction and classification problems. In such NNs, the convolutional layers (filters) extract the different features of the input images, allowing the NN to learn these filters without human intervention. The detailed schemes are shown in Supplementary Fig. S1 [54]. In the UCNN, the convolutional encoding part consists of downsampling small filter kernels that allow capturing the image features. It is followed by a deconvolutional decoding part, where these features are upsampled. The skip connections between the contracting and the expanding paths improve the feature extraction. The U-shaped network is followed by a convolution layer that learns to assemble an output based on the provided information. For the classification problem, we employ a parallel CNN structure that consists of four parallel CNN layers, each with a different number of output channels. Such configuration allows a better feature extraction while reducing overfitting. To minimize a mean square error cost function, an Adam optimizer with a learning rate of $1 \times 10^{-3}$ is used in both DNNs.

The DNNs are implemented using the TensorFlow-based Keras Python library for deep learning and neural networks toolbox of the Wolfram Mathematica. The constructed DNNs were trained and deployed on a single NVIDIA GeForce RTX 2060 graphic processing unit.

## 2. Coupled dipole method.

To demonstrate the principle of far field subwavelength imaging, first, we perform a numerical analysis using a semi-analytical model based on two-dimensional coupled dipoles [55,56]. Such a 2D model contains all the essential physical ingredients to simulate wave propagation and scattering in locally resonant media. In the model, each dipole $p_i$ is modeled by its polarizability $\alpha_i$, which follows a Lorentzian model consistent with the optical theorem, namely $\alpha^{-1} = \omega_r^2 - \omega^2 + j\left(\frac{k^2}{4\varepsilon} + \Omega\right)$ (the energy conservation demands that $Im\,\alpha^{-1} > \frac{k^2}{4\varepsilon}$), where $k$ is the wave vector, $\varepsilon$ is the medium's relative permittivity, $\omega$ is the operating frequency, $\omega_r$ is the dipole's resonance frequency, and $\Omega \geq 0$ is a part that controls absorption losses. Dipoles located at different positions $i$ and $j$ are coupled to each other through the 2D free-space Green's functions $G_{ij}(\vec{r_i},\vec{r_j}) = -j\frac{k^2}{4\varepsilon}H_0^{(2)}(k|\vec{r_j} - \vec{r_i}|)$ and can be locally excited using an external source field $E_i^S$:

$$\alpha_i^{-1}p_i - \sum_{i \neq j} G_{ij}p_j = E_i^S.$$

In order to simulate an image source, each pixel of the image is modeled as a dipole, which is excited with an external source field $E_i^S \neq 0$ with an amplitude proportional to the intensity of such pixel.

Furthermore, we eliminate the correlation between the digit shape and the total amplitude of the field by normalizing the sum of amplitudes of each dipole to the number of active pixels. The resulting linear system is solved at each frequency using the LU decomposition with partial pivoting and row interchanging. The distributions of the field at each frequency are obtained by summing the contributions of dipoles. For the image reconstruction and recognition, the fields are calculated at four separate points, placed either in near or in the far field, depending on the considered scenario. The absorption losses in the dipoles that form the metalens are simulated by adding a non-zero inelastic part $\Omega > 0$ to the imaginary part of the inverse polarizability.

## 3. Experimental setup.

We assembled an 8 × 8 lattice of speakers [inset in Fig. 3(a)], each having a resonant frequency $\omega_r \approx 290$ Hz. The overall lateral dimensions are 268 × 268 mm (≈1/5 of minimum wavelength). In this lattice, each speaker represents one pixel of the subwavelength image, and the voltages applied to speakers fix the pixels' intensities. The speakers are connected directly to the output channels of a Speedgoat

Performance Real-Time Target Machine with IO131 interface, which is controlled by the xPC target environment of MATLAB/Simulink and allows controlling the amplitudes and phases of the applied voltages. To excite our setup we use a Gaussian-shaped pulse modulated at 250 Hz and with a pulse width of 7 ms. The far-field pressure measurements are performed using four ICP microphones that are placed at a distance of 2.5 m away from the image source and separated by an angle of ≈ 40°. The ICP microphones are connected to the same target machine, which measures the amplitude and phase of pressure at these positions and stores these values on the controlling computer.

### 4. Extracting *S*-parameters.

To characterize the resonant properties of Helmholtz resonators, we performed a simple two-port scattering experiment. We placed a Helmholtz resonator inside a tube supporting a single plane wave-like propagating mode, forming a 2-port scattering network. In this picture, each port represents the wave propagating towards (away from) the resonator on both sides of the waveguide [57]. Thus, we can define the relation between the propagating waves using the *S*-matrix:

$$\begin{pmatrix} E_m^1 \\ E_p^2 \end{pmatrix} = \begin{pmatrix} S_{11} & S_{12} \\ S_{21} & S_{22} \end{pmatrix} \begin{pmatrix} E_p^1 \\ E_m^2 \end{pmatrix},$$

where $E_p^i(E_m^i)$ are the complex amplitudes of the waves propagating towards (away from) the resonator at ports $i$=1,2. The amplitudes $E_p^i(E_m^i)$ are extracted from complex pressure measurements performed at four different points along the tube (two on each side of the resonator). By performing two independent measurements, exciting the same structure from both sides, we get enough information to find all four *S*-matrix coefficients.

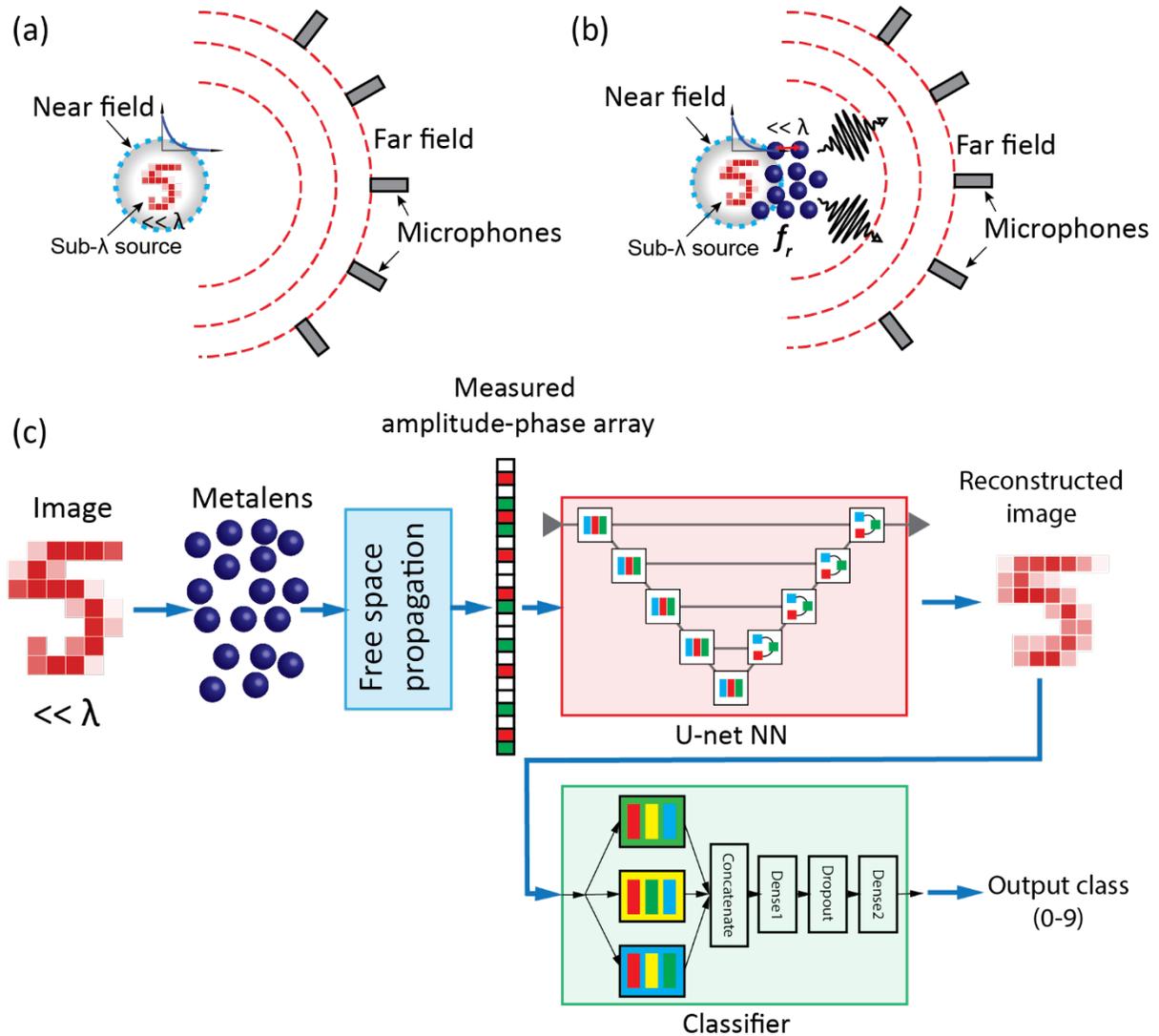

FIG. 1. Deep neural network approach for far-field recognition of subwavelength images. (a) A subwavelength source radiates into an infinite number of plane waves in all directions. The waves with high amplitude wavevector components that contain information about the subwavelength features are concentrated in the near field of the object due to their exponential decay, therefore resulting in the loss of subwavelength features in the far field. (b) A metamaterial lens inserted in the near field of the object can couple to the evanescent field components and re-radiate the waves with the information encoded into the far field patterns. (c) The UCNN (U-net convolutional neural networks) learns the correlation between the far field amplitude and phase patterns and the subwavelength images. This DNN is composed of several blocks containing a convolutional encoding front end and deconvolutional decoding back end with skip connections (see Appendix for additional details) and followed by a PCNN that classifies reconstructed images into ten categories of handwritten digits (0-9).

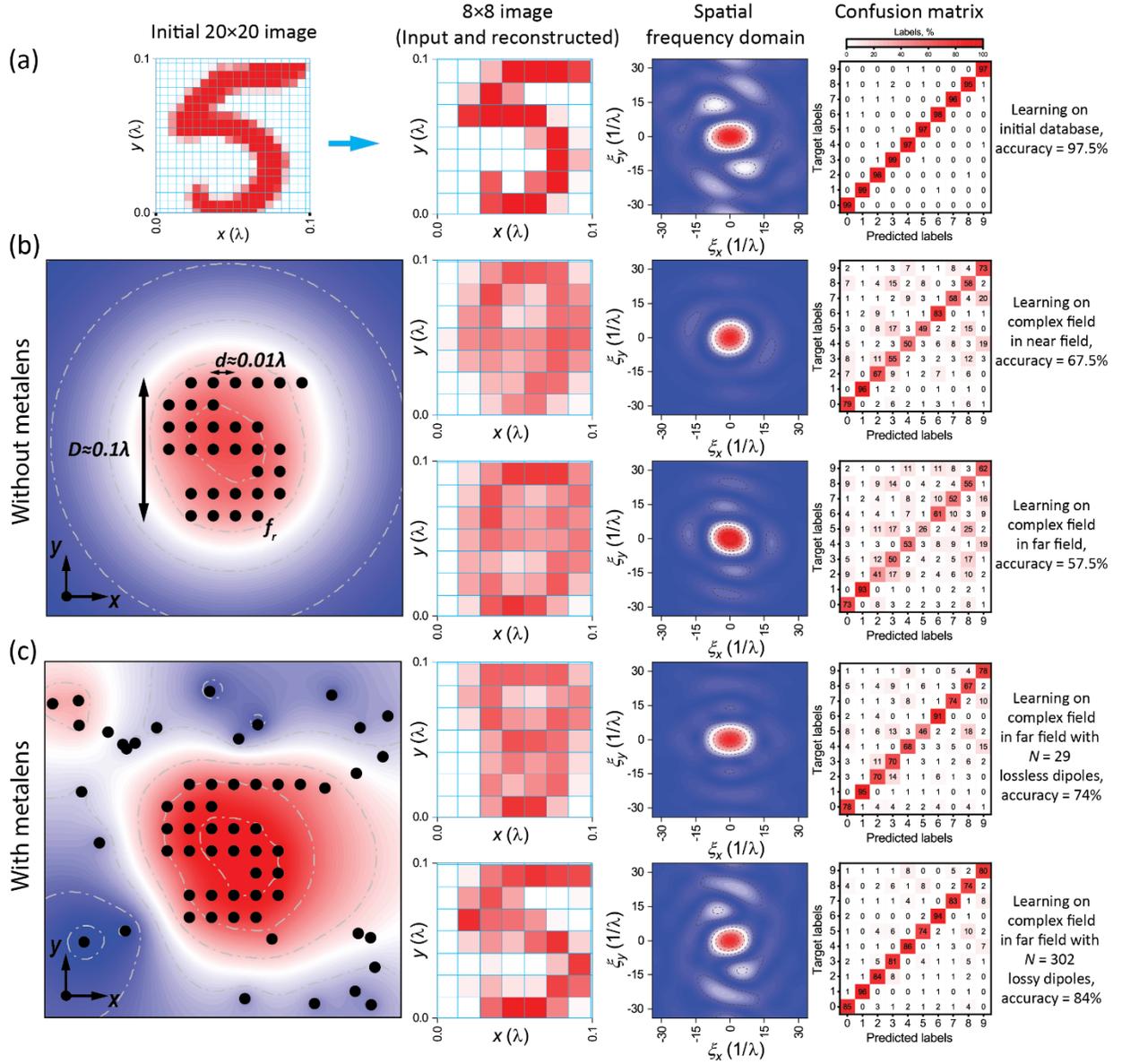

FIG. 2. Demonstration of subwavelength handwritten digits reconstruction and classification based on a semi-analytical model of 2D coupled dipoles. (a) The initial 20×20-pixel image of a handwritten digit taken from the MNIST database (left inset) and downsampled to 8×8-pixel image (second column) with corresponding distribution in spatial frequency ξ space (third column). The PCNN classifier trained on the downsampled database reaches an accuracy of 97.5% (the corresponding confusion matrix shown in the right inset). (b) The subwavelength image is modeled by an array of dipoles placed at distances d << λ and with resonance frequency $f_r$ >> $f_0$ to inhibit multiple scattering resonance. Reconstructed image (second column) and its corresponding reciprocal space domain (third column), which show the failure of UCNN to recover the higher spatial frequencies. Thus, the accuracy of PCNN classifier trained on these reconstructed images reaches ≈67.5% in the near field (top row inset) and 57.5% in far field R>>λ (bottom row inset). (c) Introducing a metamaterial lens that consists of $N$ dipoles with resonance frequency $f_r$ ≈ $f_0$ results in the creation of different multipoles that can radiate, therefore transferring the near field information to the far field. The trained UCNN manages to reconstruct better the sub-diffraction details of the images, and as a result, PCNN performs with a higher accuracy ≈ 74% for the metalens composed by $N$ = 29 lossless resonators (top row inset). The reconstruction and classification accuracy can be further improved by increasing number of dipoles ($N$ = 302) and adding losses, with the overall accuracy ≈ 84% (bottom row inset).

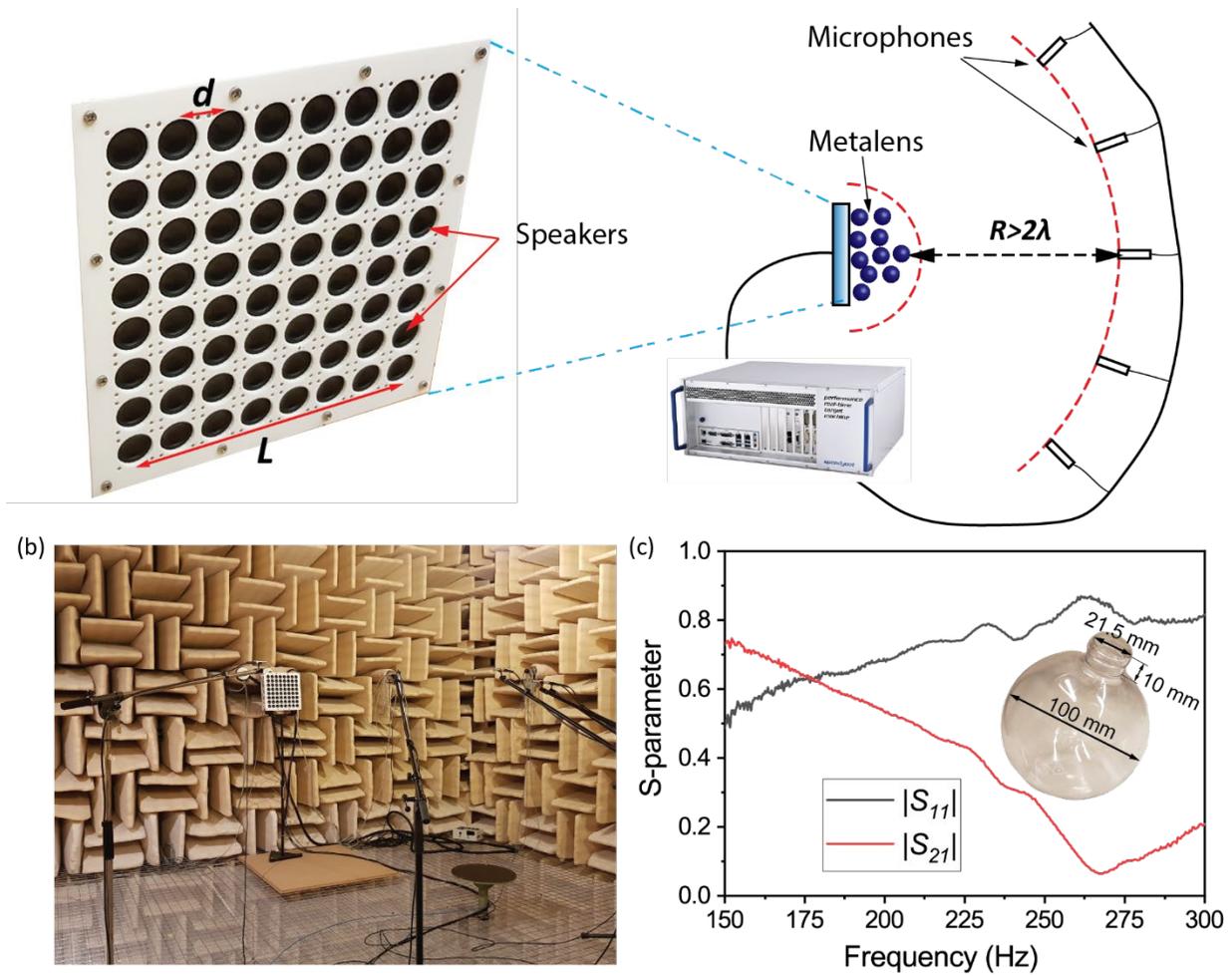

FIG. 3. Experimental testing of subwavelength acoustic imaging from far field. (a) Scheme of the experimental setup for the subwavelength image reconstruction and recognition. The acoustic image is drawn on a source (left inset), which is composed of speakers placed in a lattice with a period of $d$ = 33.3 mm (≈$\lambda_0$/41) with the overall side length of $L$ = 268 mm (≈$\lambda_0$/5), where $\lambda_0$ is the minimum measured wavelength. The amplitude of the excitation of each speaker is controlled by a Speedgoat real-time target machine, which also records the signals coming from four microphones placed in the far field. (b) Photo of the experimental setup in the anechoic chamber. (c) The scattering ($S$) parameters of a Helmholtz resonator (a plastic ball with a bottleneck as shown in the inset photo) that comprises the locally resonant metalens and has the resonance frequency $f_r$ ≈ 268 Hz.

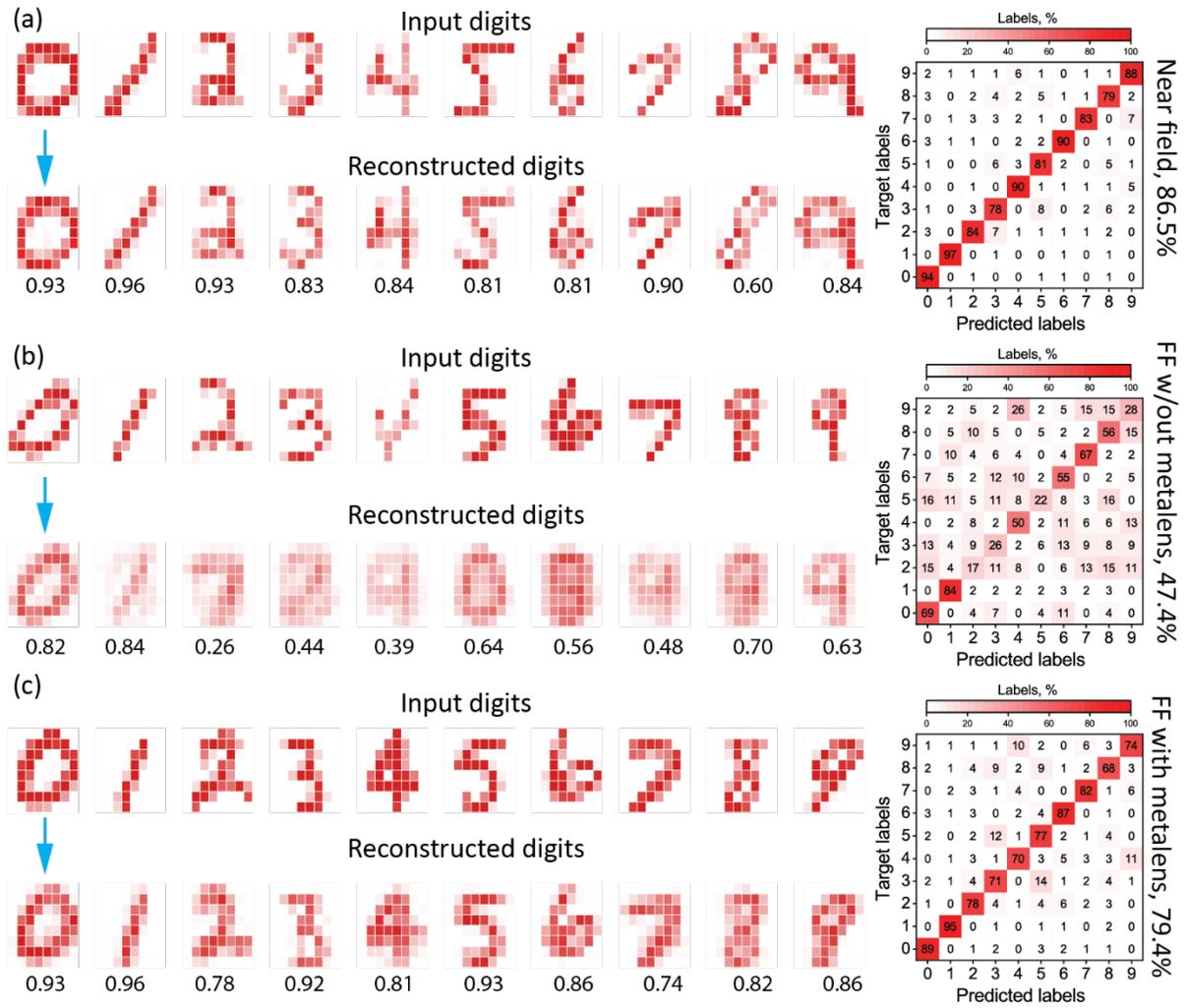

FIG. 4. Experimentally reconstructed acoustic images and their classification. Performance of the DNNs for the experimental reconstruction and classification of the handwritten input digits. (a) The reconstructed (left inset) images from the measured amplitude-phase arrays of pressure and corresponding confusion matrices for the classification problem (right inset) in (a) near field, (b) far field without any metalens and (c) far field with the metalens placed close to the source. The correlation coefficient between each reconstructed image and its corresponding target image is shown at the bottom of the image.

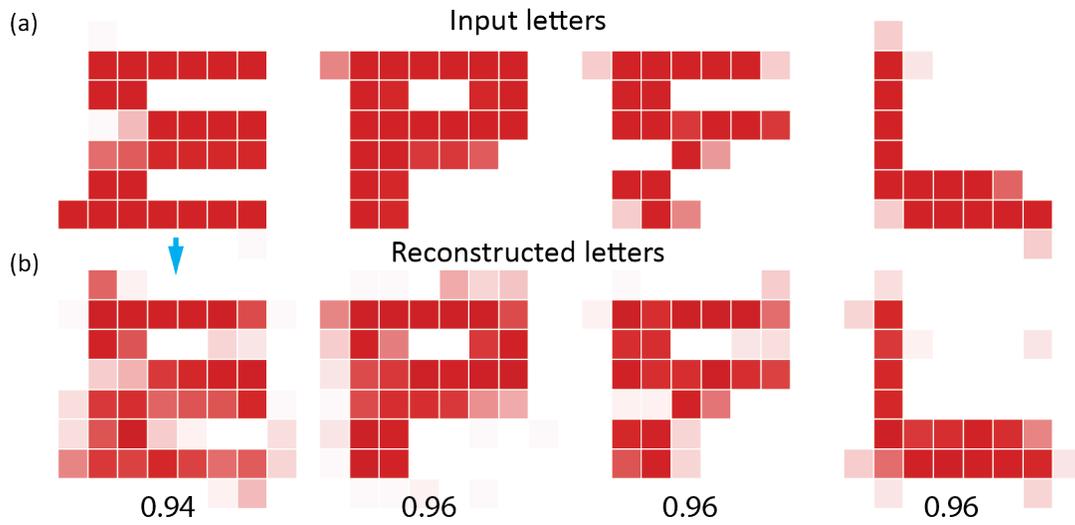

FIG. 5. Subwavelength image reconstruction using the transfer-learning technique. UCNN initially trained on the 0-9 digits is retrained on a small set (600 training and 200 test samples) of letters ("**E**," "**P**," "**F**" and "**L**") and then used to reconstruct the letters that it did not see before. (a) Initial test letters drawn on 8x8-speaker lattice. (b) Reconstructed letters using retrained UCNN, which are practically identical to the target letters, demonstrating the great versatility of the DNN approach to recognize any image using the transfer-learning technique. The correlation coefficient between each reconstructed image and its corresponding target image is shown at the bottom of the image.

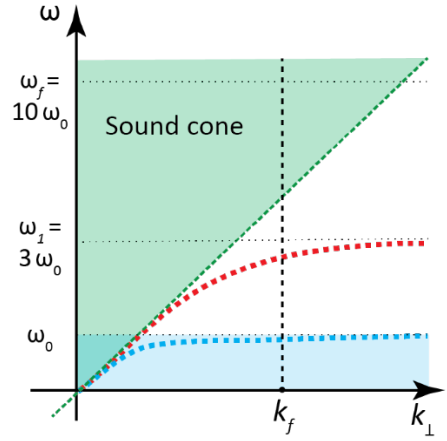

FIG. 6. Subwavelength imaging principles. Dispersion relations of resonant media composed of resonators with different resonance frequencies $\omega_0$ (blue dashed line) and $\omega_1 = 3\omega_0$ (red dashed line). Both lines lie below the sound cone (semi-transparent green region). Only the first resonant medium modes possess modes with high transverse wavevector components $k_\perp$ up to $k_f$ below $\omega_0$, i.e., in the subwavelength regime.

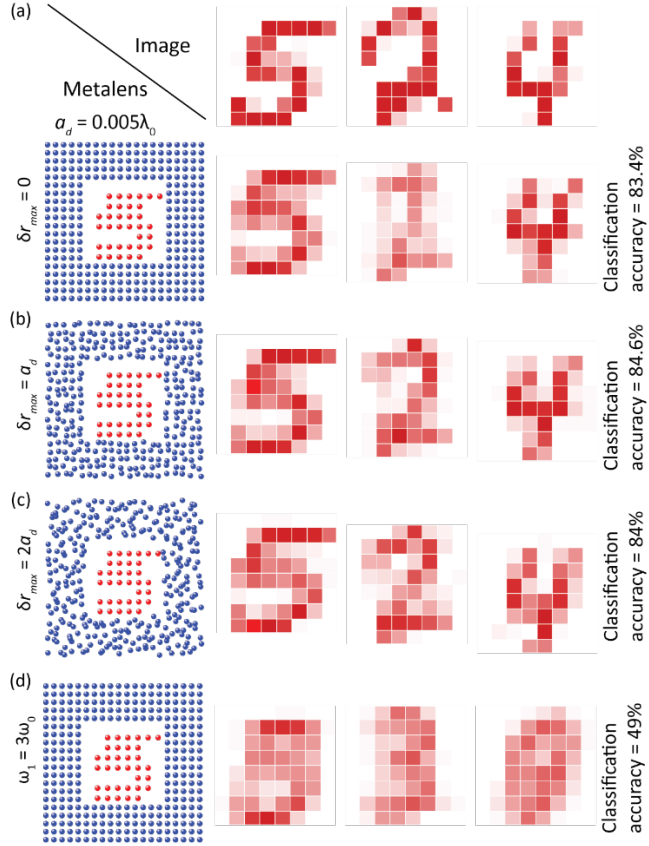

FIG. 7. Subwavelength image reconstruction using resonant metalenses. (a) A lattice of 300 resonators with a period $a_d = 0.005\lambda_0$ and a resonance frequency $\omega_0$ (left inset). The examples of input (top row) and reconstructed (second row) images using UCNN. A lattice with random displacements of resonators in the range (b) $\delta r = [0, a_d]$ and (c) $\delta r = [0, 2a_d]$ and the corresponding reconstructed images. (d) A lattice of 300 resonators with a period $a_d = 0.005\lambda_0$ and a resonance frequency $3\omega_0$ (left inset) and corresponding reconstructed images.

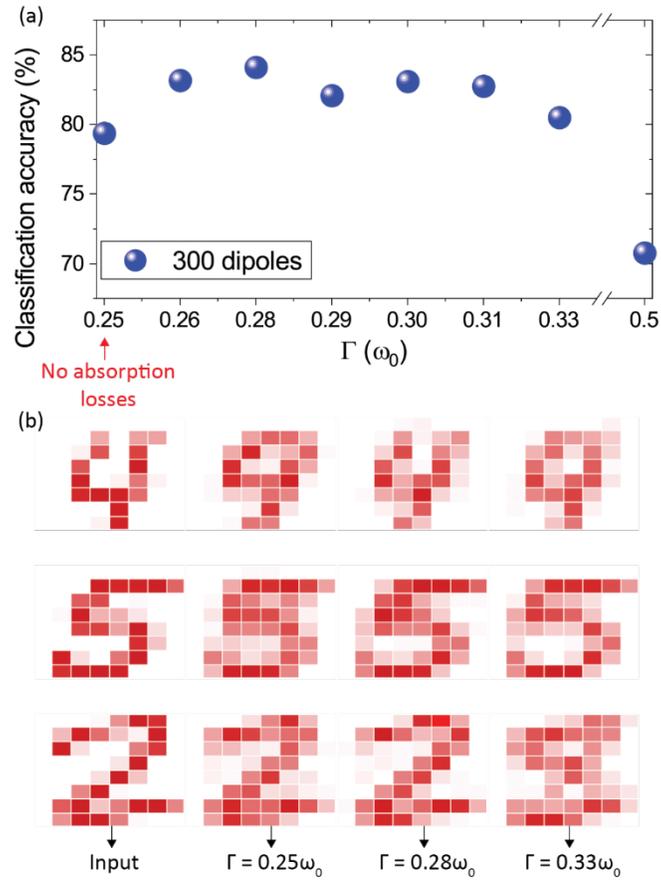

FIG. 8. Influence of absorption losses on image reconstruction using UCNN and coupled dipole method. Examples of subwavelength digits reconstructed from arrays of complex field values obtained for different collision frequencies.